\def\year{2018}\relax
\documentclass[letterpaper]{article} 
\usepackage{aaai18}  
\usepackage{times}  
\usepackage{helvet}  
\usepackage{courier}  
\usepackage{url}  
\usepackage{graphicx}  
\usepackage[usenames, dvipsnames]{color}
\begin{document}

\title{Quantifying the Impact of Cognitive Biases in Question-Answering Systems}
\author {Keith Burghardt\\
  Information Sciences Institute\\
 University of Southern California\\
keithab@isi.edu
\And
  Tad Hogg\\
  Institute for Molecular Manufacturing\\
tadhogg@yahoo.com
\And
  Kristina Lerman\\
  Information Sciences Institute\\
 University of Southern California\\
lerman@isi.edu
  }
\maketitle
\begin{abstract}

Crowdsourcing can identify high-quality solutions to problems; however, individual decisions are constrained by cognitive biases. We investigate some of these biases in an experimental model of a question-answering system. In both natural and controlled experiments, we observe a strong position bias in favor of answers appearing earlier in a list of choices. This effect is enhanced by three cognitive factors: the attention an answer receives, its perceived popularity, and cognitive load, measured by the number of choices a user has to process. While separately weak, these effects synergistically amplify position bias and decouple user choices of best answers from their intrinsic quality. We end our paper by discussing the novel ways we can apply these findings to substantially improve how high-quality answers are found in question-answering systems.
\end{abstract}

\section{Introduction}


According to the wisdom of crowds, a large group can collectively find a better solution to a problem than a typical individual~\cite{JuryThm,Galton1908,surowiecki2005wisdom,Kaniovski}.
This effect has become the foundation of crowdsourcing on the web, including systems for content creation~\cite{WikiWisdomOfCrowds}, product review~\cite{lim2015evaluating}, peer recommendation~\cite{PopularDynam,SocialInfluenceBias2}, and question-answering (Q\&A)~\cite{BestAnswerYahoo,HighQualityQA}. In many cases, the crowd's solution aggregates many users' recommendations or votes as they are sequentially added. Recent work has suggested that this method should optimally determine the best items \cite{Celis2016,Krafft2016}, and displaying item popularity is a simple way to make high-quality items easier to find. However, individual decisions can be affected by cognitive biases, which may compound to decouple the relation between wisdom (the quality of ideas) and crowds (popularity).

For example, recent research has demonstrated that social influence introduces correlations between decision makers, which can reduce the quality of collective solutions~\cite{Lorenz2011,Kaniovski} and make them less predictable~\cite{SocialInfluenceBias,SocialInfluenceBias2,MusicLab}.
Empirical studies of crowdsourcing systems suggest that users' bounded rationality, and reliance on heuristics like item position ({\it position bias}), are even more important limiting factors in collective performance~\cite{PopularDynam,MyopiaCrowd}.

\paragraph{Our Contribution}
In this paper, we examine cognitive factors that affect crowd performance in order to understand how to enhance the wisdom of crowds, and better correlate item popularity with ``quality''. We define quality as properties of the answer's text, such as how well the answer addresses the question or how well it is written, which are independent of where or how the answer is shown to users. Furthermore, we explore these factors in the context of Q\&A, a popular crowdsourcing task, because we can compare a natural experiment and empirical data to novel controlled experiments. The questions we address are:

\vspace{3pt}
\noindent
{\it Q1: What cognitive factors contribute to answer popularity?}

\vspace{3pt}
\noindent
{\it Q2: How does popularity relate to quality?}

\vspace{3pt}
\noindent
We explore these questions with two complementary approaches: (1) simulating Q\&A with experiments in which parameters are precisely controlled, and (2) using empirical data to compare our experimental results to real-world systems. The experiment allows us to carefully tease out why answers become popular without confounding variables, while comparing experimental and empirical data allows for us to check the ecological validity of the experiment. Specifically, we create an experimental model of Stack Exchange (SE), a popular Q\&A platform, using Amazon Mechanical Turk (MT). We use actual questions and answers from the \emph{English Language Learners} forum of SE, and ask MT workers to pick the best answer to each question. Our model allows us to replicate some of the functionality of SE in a controlled setting, such as the number of answers and the scores MT workers see for each answer. We also record where the workers move their mouse, a proxy for attention that agrees well with eye-tracking data~\cite{MouseEye}. 


In addition, we use data from SE to analyze a natural experiment in which users vote for answers that are ordered in different ways while controlling for perceived popularity. The major takeaway from the empirical data is that increasing the answer position even slightly can increase the probability it is chosen {\it by up to 30\%} compared to ordering answers at random, and that this increase is in large part aided by cognitive load. We further find qualitative agreement between experimental data (where answers are ordered arbitrarily) and empirical data (where answers are presumably ordered by their quality, if popularity and quality are correlated) on the probability of voting for answers versus the answer position. This agreement suggests that users choose answers in large part because of their position on the webpage and not due to their quality.

The rest of the paper is organized as follows. We first discuss the background of crowd wisdom and cognitive biases in user behavior. Next, we describe our MT experiment and the natural experiment in SE, and then show our results in greater detail. We finish by discussing the positive and negative consequences of position bias in Q\&A systems, and ways in which crowdsourcing can be improved.

\section{Related Work}
The wisdom of crowds, in which a large group, even one composed of uninformed individuals, can collectively reach a better decision than individual experts, was first theoretically predicted in a study of juries~\cite{JuryThm}. When jurors are homogenous and their decisions uncorrelated, the majority decision of a jury is almost always of higher quality than any individual juror's. This presaged the work by Galton and others~\cite{Galton1908,surowiecki2005wisdom}, in which they empirically found that the mean of many individual guesses was a better prediction than any expert, because the mean averages out uncorrelated errors. In recent years, wisdom of crowds was reapplied to several different fields, such as crowd sourced information~\cite{WikiWisdomOfCrowds} and Q\&A forums~\cite{BestAnswerYahoo,HighQualityQA}, prediction markets~\cite{PredictMarket}, and peer recommendation~\cite{PopularDynam,SocialInfluenceBias2}. Importantly, these applications rely on the assumption that an item's popularity indicates its quality, which, although not necessarily a corollary of early work on crowd wisdom, is sometimes a reasonable assumption~\cite{PerformanceSuccess}.

Recent interest, however, has focused on how item popularity can enhance or reduce the wisdom of crowds. Sequential voting, which is commonly used in crowdsourcing systems \cite{PopularDynam,SocialInfluenceBias,SocialInfluenceBias2,BestAnswerYahoo,HighQualityQA} can theoretically improve collective quality compared to non-sequential votes~\cite{Celis2016,Krafft2016}. However, Lorenz et al.~\shortcite{Lorenz2011} found that when users knew the guesses of other users, the variance of guesses drops dramatically, and there is increased confidence in guesses that still differed significantly from the correct answer. Similarly, theoretical work suggests that correlated opinions can reduce the quality of collective guesses \cite{Kaniovski}. Salganik, Dodds, and Watts~\shortcite{MusicLab} also found that social influence can affect which items become popular independent of intrinsic factors of an item. However, more recent models of the same data suggest that answer position was an underappreciated \cite{MusicLabModel}, and potentially much more important factor to explain why an item was chosen \cite{PopularDynam}.

Position bias is an effect in which items listed first are more likely to be chosen. This effect occurs, for example, in voter ballots~\cite{BallotOrderCA,BallotOrderCA2}, search engines \cite{AttentionClick,AttentionClick2}, information aggregation sites~\cite{PopularDynam}, and peer evaluation~\cite{lerman14as}. Although the effect is usually observed when people choose among many items, only recently have researchers explored how this effect appears when there are few items to choose from \cite{MyopiaCrowd}. In addition, mechanisms underlying position bias have not been fully characterized. The primacy effect, in which items seen first are more likely to be chosen~\cite{Primacy1}, may play a role, but other effects such as attention~\cite{DDM3Objects}, trust~\cite{AttentionClick}, or popularity could contribute to position bias as well. One goal of this paper is to decouple these factors in order to determine what causes position bias in real systems.

Similar to some previous work \cite{HighQualityQA,BestAnswerYahoo}, our MT experiment tests crowd wisdom by exploring why certain answers are upvoted or accepted by askers in Q\&A boards. A recent paper suggested that users appear to vote or accept answers in Q\&A boards for reasons other than the intrinsic qualities of an item, especially when there are many answers to choose from \cite{MyopiaCrowd}. Our paper improves upon this work by disentangling position bias from popularity and quality, and exploring the many cognitive factors that can increase position bias.

We will show the number of answers users see increases position bias independent of answer quality. A plausible reason for this effect is because users can only processes a limited amount of information. When more information is visible, users decrease the effort they are willing to spend evaluating each piece of information due to information overload~\cite{InfoOverloadPsychology,TwitchInfoOverload}. Finally, we use natural and controlled experiments because empirical data alone could be affected by correlations between attributes, while experiments alone may not capture all relevant aspects of real-world scenarios. Agreement between experiments and empirical data increases confidence in the results, and separately validates each approach~\cite{ObservationalVsExp}.

\section{Methods}

We used both a controlled experiment with MT workers and a natural experiment resulting from a change in the SE platform.

\subsection{Mechanical Turk Experiment}
\begin{figure}[t]
	\centering
	\includegraphics[width=0.9\columnwidth]{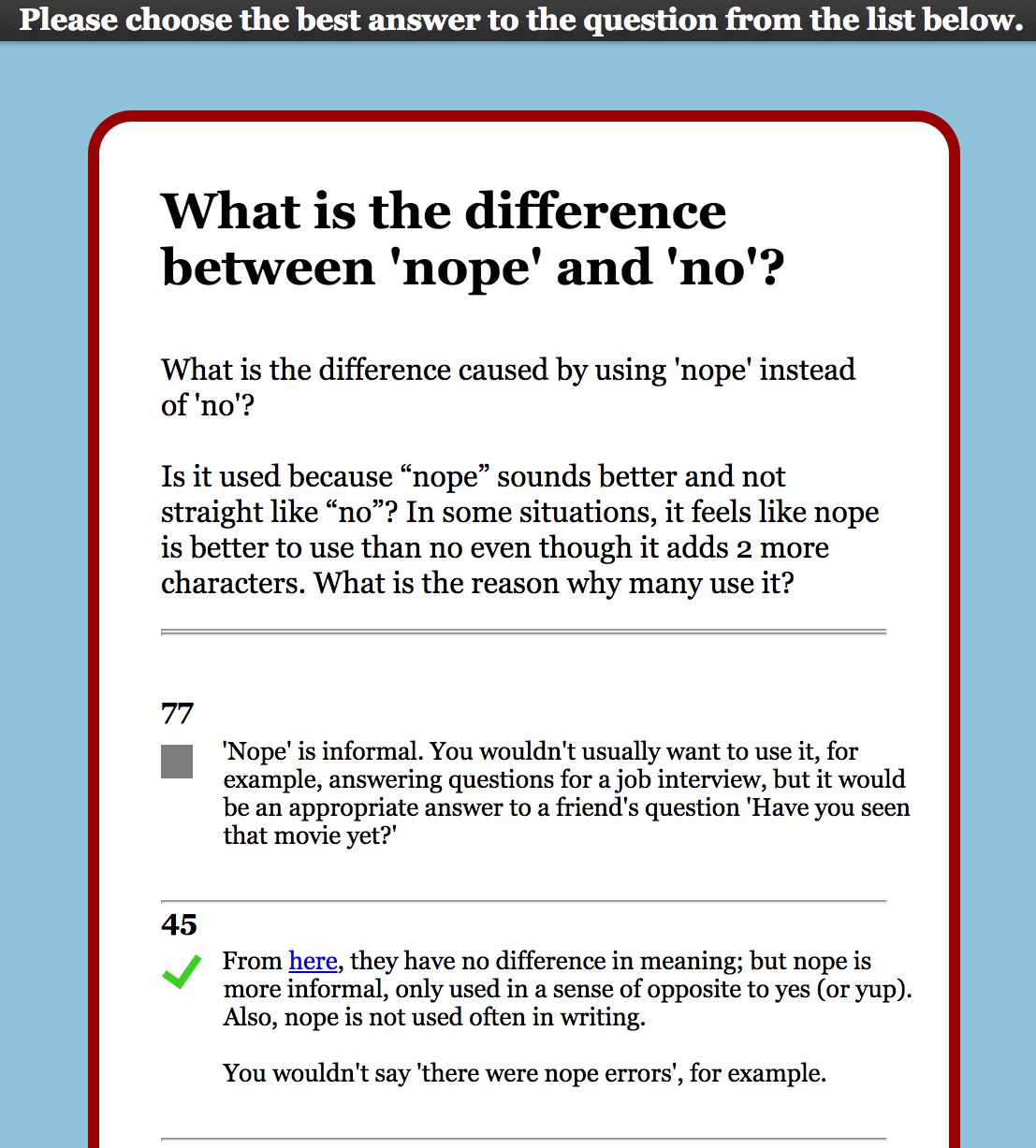}
	\caption{\label{fig:ExampleOutput}An example screen used in our experiment, showing a question from SE (http://ell.stackexchange.com/questions/30/what-is-the-difference-between-nope-and-no). In this screenshot, the number of votes are visible, representing the social influence experimental condition (in the random experimental condition, the numbers of votes are hidden) and a checkmark is next to the chosen answer. After an answer is chosen, users click an ``accept'' button to progress to the next question.
	}
\end{figure}

Our experimental model of Q\&A directs registered Amazon Mechanical Turk workers to a web page instructing them to ``choose the most correct answer for each of ten questions" (see Fig. \ref{fig:ExampleOutput} as an example). The page models Stack Exchange, specifically the \emph{English Language Learners} (ELL) forum, from which the questions were selected. Each of the questions had at least 8 answers. The workers are mostly from the US, Canada, or Britain ($90\%$ among workers sampled, based on IP addresses), where English is commonly spoken. Our choice of the ELL forum is meant to increase the similarity between workers and SE users, because the questions and answers were meant to be accessible for both native and non-native English speakers. We only show ten questions to each user because we expect the quality of workers' choices to decline appreciably if they are asked a large number of questions, according to recent research on performance depletion in Q\&A systems~\cite{Ferrara17}. Further, we limit the total number of answers workers could see and vote on in order to create sufficient statistics on the popularity of each answer.

In the experimental model, each MT worker is assigned to one of two experimental conditions. In the first (``random'') condition, workers see answers listed in a random order below the question (independently for each worker) and no score is shown. In the second (``social influence'') condition, the scores are shown next to each answer, and answers are ordered by score. In both cases, the 2 or 8 oldest answers from the ELL website were listed below their associated question, which is consistent with the answers SE users would have seen.

The workers assigned to the social influence condition are told that ``scores listed next to each answer denote the number of individuals who chose this answer in the past'' (as on SE). However, in reality, the ``scores'' are independent and randomly generated numbers from 0 to 100 in the 2-answer scenario and from 0 to 25 in the 8-answer scenario (such that the scores add up to 100 on average). We generate these numbers independently for each worker. In short, answers are ordered randomly in both experimental conditions, but in one, workers think that other workers upvoted particular answers.

We recruited workers with an approval rate of over $95\%$ and more than 1000 Human Intelligence Tasks (HITs) completed. For Trial 1 in the random experimental condition (shown in Table~\ref{tab:summary}), we requested ``Masters" workers, i.e., people Amazon labels as especially high performing. Their voting behavior was statistically similar to that of other users. Because it takes orders of magnitude more time to find enough workers who are Masters, we dropped this requirement in later experiments. Workers were given up to one hour to complete an assignment (the median time is $8.0$ minutes in the random condition, and $9.5$ minutes in the social influence condition). Each worker was paid $\$0.50$ for completing the assignment. The equivalent hourly wage is half that originally designed because the tasks took unexpectedly long compared to initial tests in which the authors were subjects.

\begin{table}[t]
\centering
\caption{Number of questions in each experiment trial.*}
\label{tab:summary}
\begin{tabular}{|c|c|c|c|}
  \hline
\vspace{-9pt}&&&\\
\# Answers &Trial  & \# Questions  & \# Questions  	\\
	&  & (Random) &  (Social Influence) \\\hline
\vspace{-9pt}&&&\\
2 Answers&  Trial 1 &440  & 438 \\
&  Trial 2 &473  & 174 \\
\hline
\vspace{-9pt}&&&\\
8 Answers&  Trial 1 & 410 & 412 \\
&  Trial 2 & 930 & 1256 \\
&  Trial 3 & 447 & --\\
  \hline
\end{tabular}
\begin{flushleft}
*270 \& 228 workers are in the random and social influence conditions, respectively.
\end{flushleft}
\end{table}

Once workers choose an answer, they click a button to advance to the next question. After completing the last question, workers are given a six-digit ID, which they submit in order for us to associate a particular worker with a given experiment. Experiments that are retaken by workers, or do not have an associated ID, are removed from analysis. Occasionally, so many people submit their completed task to our server that lines of data overlapped, making analysis of this raw data more difficult, therefore data in these cases are discarded. In the random experimental condition, we have clean data from $270$ workers out of $323$ completed experiments, and in the social influence experimental condition, we have clean data from $228$ workers out of $250$ completed experiments.
We perform multiple trials for each condition, as listed in Table~\ref{tab:summary}.
In the experiments, we record:
\begin{enumerate}
\item the question number
\item the number of answers for each question
\item the time a worker answers each question
\item the answer a worker chooses
\item the order answers are listed for each question
\item the times when workers scroll their computer mouse (or track pad) over an answer
\item each answer's score (if applicable), and
\item the start and end time for a worker to complete all questions
\end{enumerate}

In the first and last trials in the random condition, and trial 1 in the social influence condition (see Table~\ref{tab:summary}), the number of answers are randomly chosen to be either 2 or 8 for each question. Trial 2 in both conditions always has 8 answers. We check the consistency of behavior in these trials by comparing the probability a worker chooses an answer versus its position between trials 1 \& 2 using the Kolmogorov-Smirnov (KS) test. This test shows no statistically significant differences between the distributions (p-values $>0.1$). We further compare the popularity of answers (averaging over the answer position) across the two worlds and find high correlations ($\rho = 0.67$ \& $0.80$ for 2 and 8 answers visible, respectively), therefore separate trials and experiments produce consistent results.

\subsection{Natural Experiment}

In August 2009, SE changed how it ordered answers with the same score from chronological (oldest to newest) to random order \cite{Oktay10}. This change forms a natural experiment for how answer ordering affects users’ choices. To exploit this change, we select an appropriate part of SE and create a dataset as follows.

We look at all votes on all technical and meta boards on SE from August 2008 until September 2014\footnote{\textit{https://archive.org/details/stackexchange}}. Boards labeled as ``technical'' by SE typically cover programming questions. Meta boards provide a forum to discuss a specific board. For example, ``Meta Stack Overflow'' discusses topics relating to the board ``Stack Overflow''. We split the data to control for Simpson's paradox, in which behavior seen in aggregated data can differ significantly from the disaggregated data~\cite{SimpsonEffect}. We do not analyze votes where (1) more than 2 answers have the same score as the answer voted on, and (2) there is an accepted answer, which can affect answer positions and may provide an additional social signal to a voter. Finally, we split data by the number of answers users would see when they cast their vote, based on the day they voted. The way in which we reconstruct the score for each answer, in order to determine which answers have the same score, is discussed below. In preliminary work, we further divided the data by the position the 2 answers appeared in (the top or bottom of the page), and when the votes occurred (6 months before and after the change, or the entire time period), but this does not affect our conclusions, therefore we re-aggregated the data, and only focus on the relative position of the 2 answers. We focus on the Stack Overflow data because it is the largest board (with 200K votes in the months before, and 2.8M votes in the months after, the rule change), but we find qualitatively similar behavior in other technical boards or meta boards aggregated together.

The data we have on votes tell us the millisecond when each answer was made, the day each vote was made, the order of votes, which answer was voted on, and whether it was an upvote or downvote. From the date a vote was cast, we can determine how many answers each user saw (assuming the vote was made at the end of the day), while sequentially adding votes to the appropriate answer allows us to determine the score for each answer just before a vote was cast. For the natural experiment, we use this data to determine the order of answers with the same score, and the number of answers seen, just before a vote was cast. This allows us to better understand how answer position and cognitive load (i.e., the number of answers seen) affect voter decisions. Similarly, by knowing the order of answers with different scores, we can better understand how score, coupled with position, increases position bias. In the latter case, we focus on the probability to vote on an answer after August 2009, and all votes are made before an answer is accepted, therefore answers with the same score are ordered at random relative to each other.

\section{Results}

In this section, we discuss the factors affecting answer popularity. We will show that, controlling for answer position, the effect of perceived popularity is negligible. That said, position bias is affected by the
\begin{enumerate}
\item number of answers a user sees (cognitive load)
\item score next to each answer (perceived popularity)
\item attention an answer receives.
\end{enumerate}
Separately, these factors are small, but together they help explain position bias we see in real-world data. A natural experiment in which answers with the same score are first ordered chronologically and then at random allows us to determine position bias when controlling for perceived popularity. Furthermore, position bias for experiments in which scores are generated randomly is in surprisingly close agreement with empirical data in which answers are supposedly upvoted due to quality. Together these results suggest that position bias can strongly decouple answer quality from answer popularity.

\subsection{Origins of the Position Bias}

\begin{figure}[t]
\centering
\includegraphics[width=0.9\columnwidth]{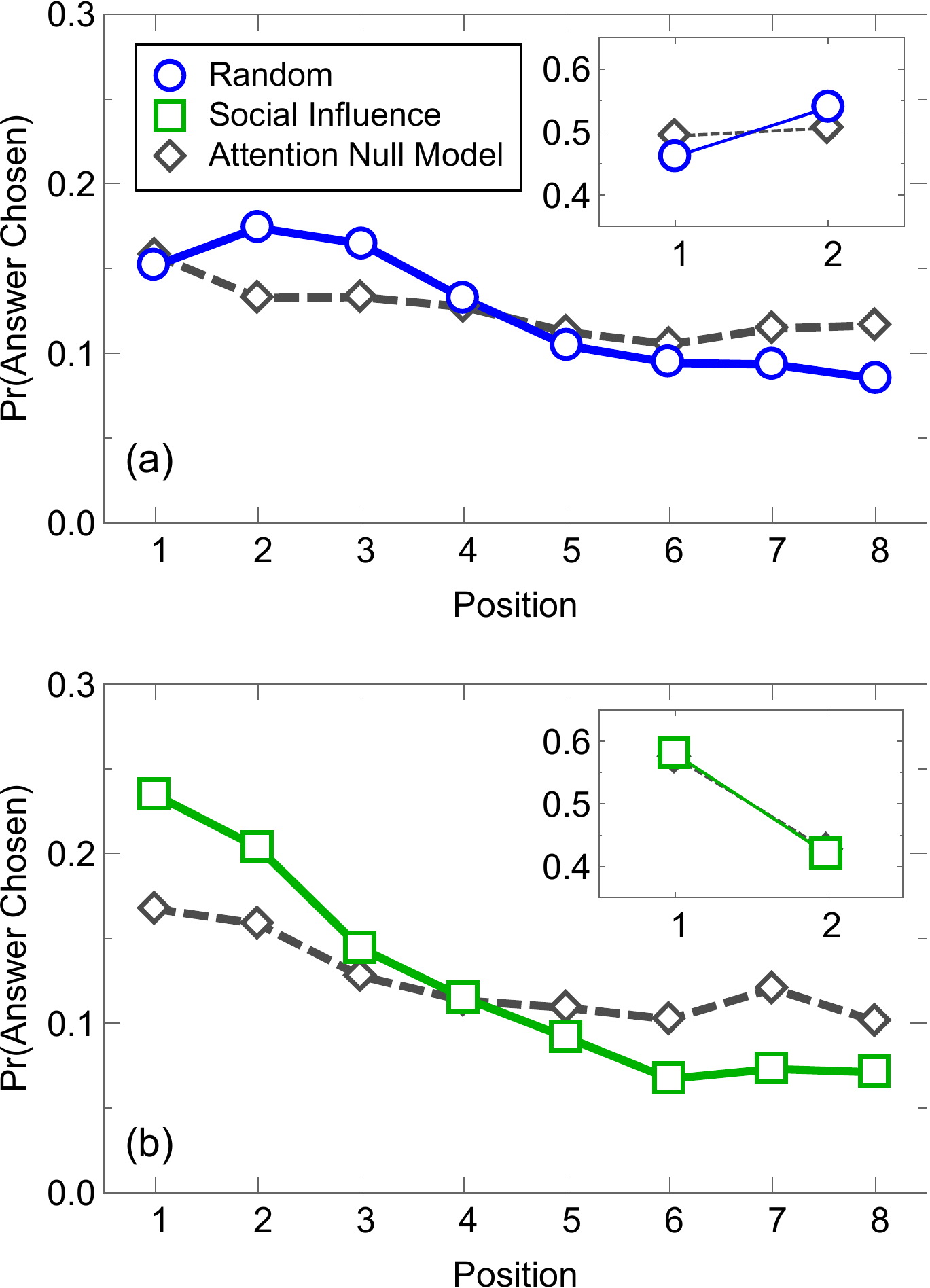}
\caption{{Probability to choose an answer versus its position in the experiment}. (a) 8 answers visible in the random condition (inset: 2 answers visible) and (b) 8 answers are visible in the social influence condition (inset: 2 answers visible). Also shown is the null attention model, discussed in the main text. Error bars for all data are smaller than the plot markers.
	}
\label{fig:PChooseAnswer}
\end{figure}

Our experiment disentangles some of the factors contributing to position bias, such as information or cognitive load, score, and attention, and how these factors affect worker decisions. Figure~\ref{fig:PChooseAnswer} shows the experimental probability a worker chooses an answer as the best answer as a function of answer's position in the list of answers under the random experimental condition (solid blue lines), the social influence condition (solid green lines), and a null model (described below) where users choose answers based on the amount of attention they receive. Main figures report conditions where 8 answers are shown, while insets report cases when 2 answers are shown. To allow for the best agreement between the null attention model and data, we removed cases in which users chose an answer that was moused over less than an arbitrary threshold of 0.01 seconds ($0~(0\%)$, $706~(40\%)$, $5~(0.008\%)$, and $641~(38\%)$ of votes were removed from Figs. \ref{fig:PChooseAnswer}a--b, respectively). The trends shown in the figures are the same when including all votes in the dataset, and when the threshold is larger, such as 0.1 seconds.

When 2 answers are shown in the random condition (Fig~\ref{fig:PChooseAnswer}a inset), workers are less likely to choose the first answer than the last one (p-value$<10^{-6}$), while when 8 answers are shown, they prefer top answers to those shown in lower positions (Fig.~\ref{fig:PChooseAnswer}a). Future work is necessary to understand why the last answer was more likely to be chosen when 2 answers were shown. This finding is not affected by including data where the chosen answer is not moused over. That said, the trend we see in which answers appearing earlier in a list are preferred as the number of answers increases is in agreement with previous research~\cite{MyopiaCrowd}. In the social influence experimental condition, on the other hand, workers are more likely to choose answers in top positions when both 2 and 8 answers are shown (Figs.~\ref{fig:PChooseAnswer}b). Just as the case when scores were not shown, the top half of the answers are more likely to be chosen as the number of answers increases (58\% and 68\% for 2 and 8 answers, respectively), although the overall probability to choose top answers increases significantly when scores are shown. Interestingly, when we control for position, there is no statistically significant correlation between score and the probability an answer is chosen (all p-values are greater than $0.1$), so {scores amplify position bias}.

We determine how attention contributes to position bias by using mouse movement, which correlates with eye tracking \cite{MouseEye}. 
Specifically, we only record when the mouse is moving over or clicking on the answer, rather than when users scroll over it with their scroll wheel, because we want to have greater confidence that mouse movements were intentional. Although mouse tracking data is not perfect, we believe it is a practical way to measure attention. To check this, we compared the probability to choose an answer versus the fraction of time a worker mouses over it, which we call \emph{time share}. We find that the larger the time share, the greater the probability a worker will choose it (Cragg \& Uhler's Pseudo $R^2$ values are between $0.44-0.54$ using logistic regression), in qualitative agreement with previous research~\cite{DDM2Objects,DDM3Objects}, in which users were more likely to pick answers that received more attention.

We create a null model in which users choose an answer due to the amount of attention it receives. In this model, the probability a user chooses an answer is directly proportional to the share of time an answer is moused over. The dashed lines in Figure~\ref{fig:PChooseAnswer} compare this null model to experimental data. We find that position bias is much stronger than the null model when 8 answers are visible (they differ significantly, p-values $<0.05$), therefore position bias cannot be fully explained by this model, but it does appear to partly explain position bias when scores are visible.

These observations lead us to the following conclusions.
\begin{enumerate}
    \item \emph{Cognitive load (number of answers visible) increases position bias}, 
  \item \emph{Perceived popularity increases the position bias},
\item \emph{Perceived popularity, when corrected for position, is not a significant factor}, and
  \item \emph{Attention increases the position bias}.
\end{enumerate}

\begin{figure}[t]
	\centering
	\includegraphics[width=0.95\columnwidth]{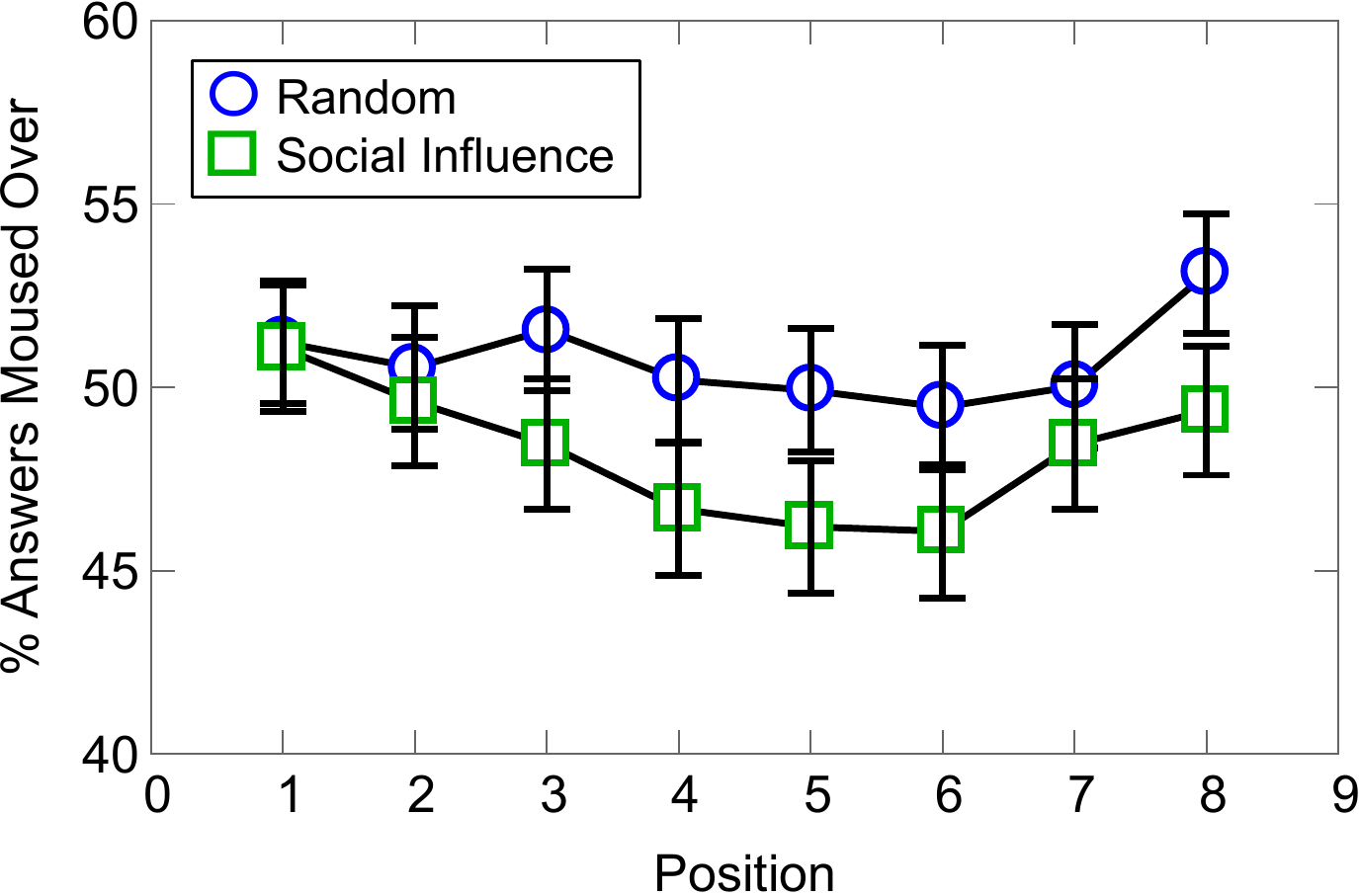}
	\caption{\label{fig:PickVsTimeShare}{The role of attention in workers' choices of answers.}
The mean percentage of times an answer is moused over versus position in the random and social influence experimental conditions. Error bars are standard errors.
	}
\end{figure}

To better understand the last point, we observe percentage of answers moused over 
versus its position (Fig. \ref{fig:PickVsTimeShare}). We find that, although the top answer is moused over with equal regularity in both the score and no-score conditions, users mouse over later answers 5\% less on average when scores are visible than when scores are not (p-values $<0.002$). Top answers therefore receive more attention, which subsequently increases the probability an answer is picked.

\subsection{Comparison with Empirical Data}

To confirm that our results are not specific to the design of the experiment, or traits specific to MT workers, we perform comparative analysis with data from SE. We used anonymized data representing all questions, answers, and votes from August 2008 until September 2014\footnote{\textit{https://archive.org/details/stackexchange}}. Results of empirical analysis, including the natural experiment, are strongly consistent with the experiment, giving confidence about its \emph{ecological validity}.

\begin{figure}[t]
	\centering
	\includegraphics[width=0.95\columnwidth]{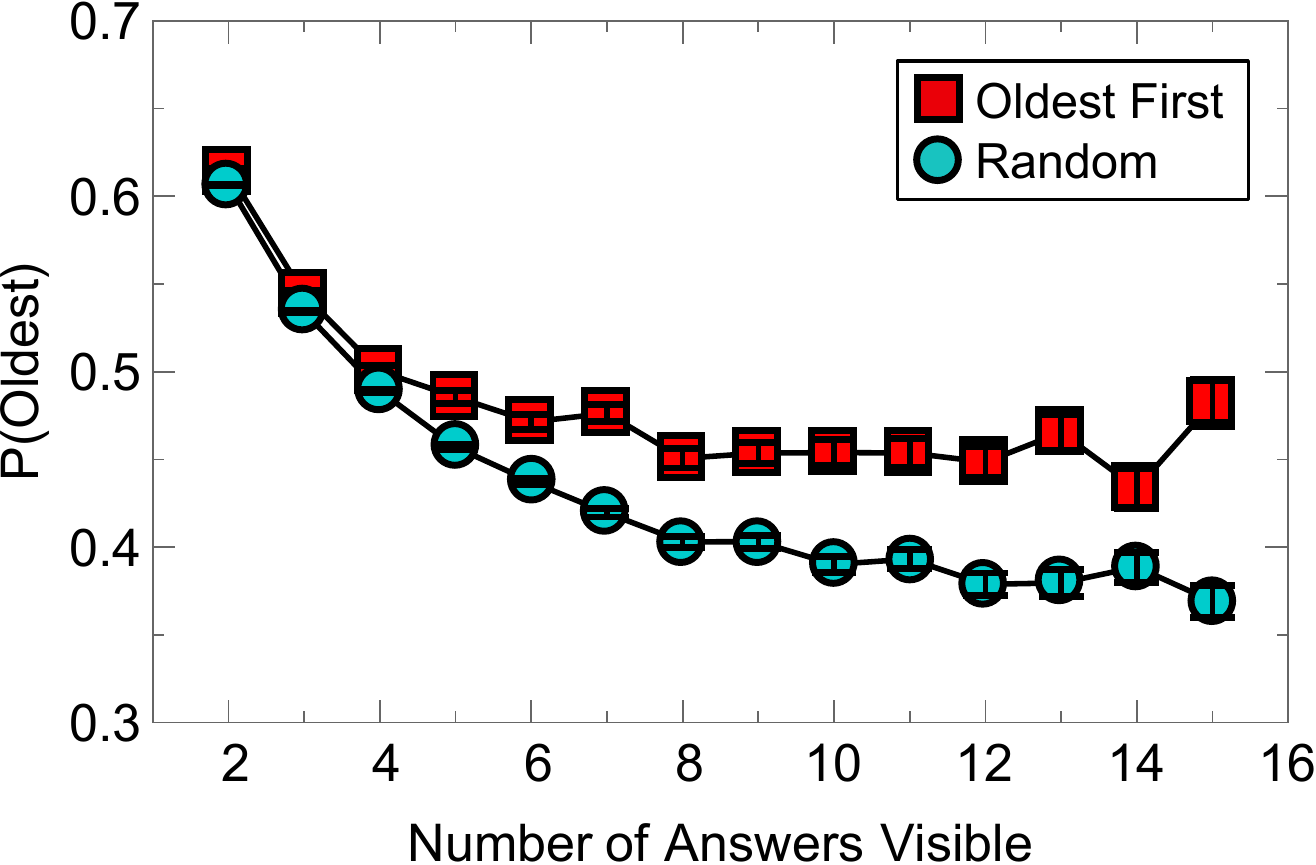}
	\caption{\label{fig:NaturalExperiment} {Position bias versus number of answers visible in a natural experiment.} The probability to vote for the older of 2 answers with the same score on Stack Overflow when the oldest answer is shown first (red squares) or at random (cyan circles). 
	}
\end{figure}

\subsubsection{Position Bias: Evidence from a Natural Experiment}

In August 2009, SE changed how it ordered answers with the same score from chronological to random order. This change allows us to test how answer ordering affects users' choices, which we plot in Fig. \ref{fig:NaturalExperiment}. We first notice that older answers are more likely to be chosen than newer answers when there are 2 or 3 answers visible, but the preference is towards newer answers as the number of answers increases. There's also an increasing preference towards the answer listed first as the number of answers increases. In fact, an older answer is up to 30\% more likely to be chosen when listed first than when answer positions are randomized (p-value$<0.001$ for all plot markers shown). 
In comparison, a previous study of this natural experiment did not find that position bias was statistically significant when all data was apparently aggregated together \cite{Oktay10}.

\begin{figure}[t]
\centering
\includegraphics[width=0.95\columnwidth]{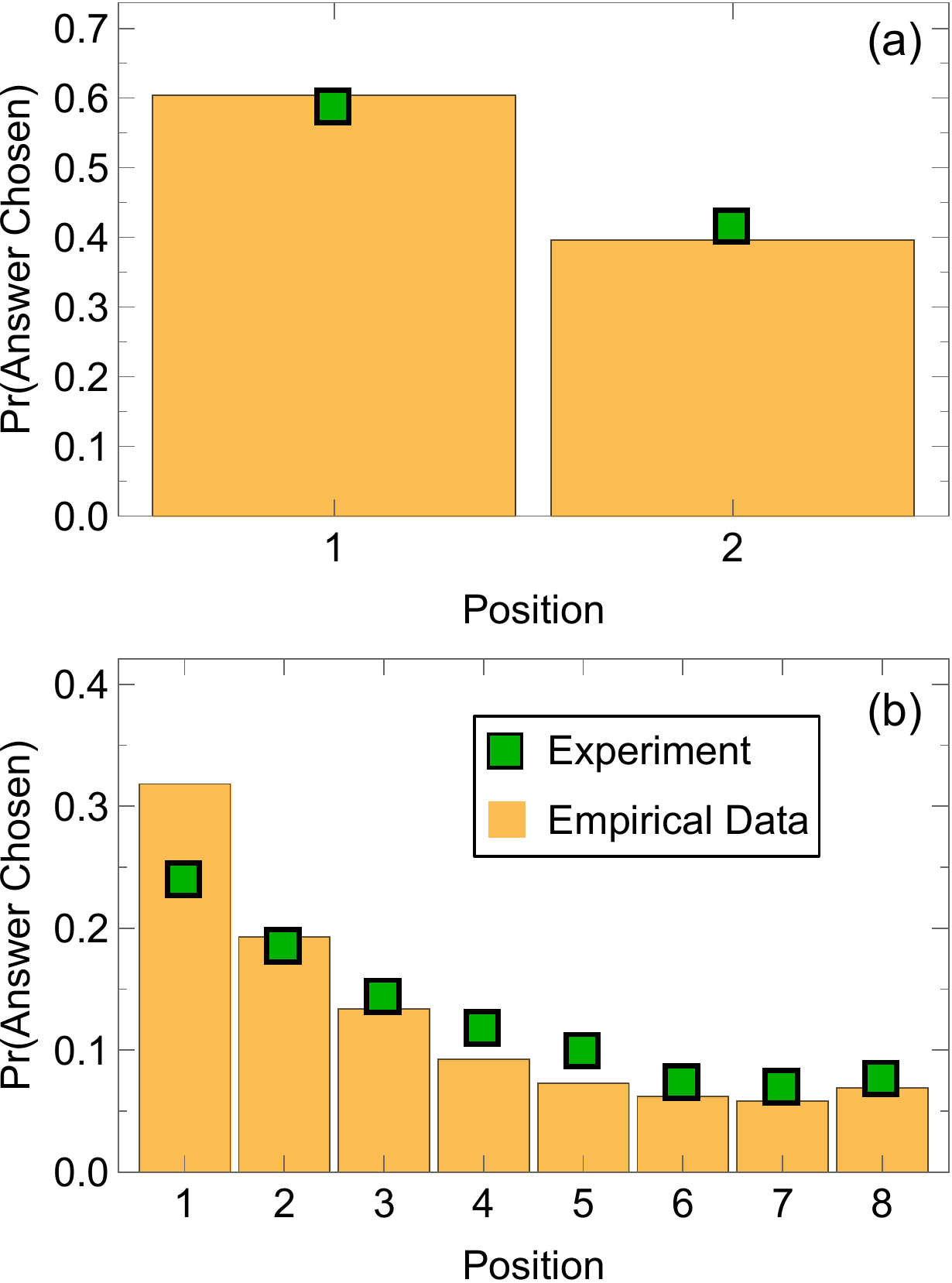}
\caption{{Comparison between experiment and empirical data.} The probability SE users in non-technical boards (yellow bars) and MT workers (green squares) vote for an answer when scores are visible and (a) 2 or (b) 8 answers are visible. Error bars are smaller than the plot markers.
	}
\label{fig:PrChoose28AnswersVoteData}
\end{figure}

\subsubsection{Perceived Popularity: Evidence from Empirical Data}

Although the previous findings suggest that answer position strongly affects the probability to choose an answer, it does not address the effect of perceived popularity. We therefore used the vote data from all non-technical SE forums from August 2009 through September 2014 to determine the aggregate fraction of votes users give to answers versus their position just before each answer has been voted on (Fig.~\ref{fig:PrChoose28AnswersVoteData}). For non-technical forums, this produces 790K votes when 2 answers are visible and 43K votes when 8 answers are visible. We see similar behavior in data from other forums, e.g., Stack Overflow. The ELL forum alone had too little data to make an adequate comparison. Answers are ordered, by default, from highest to lowest score, which provides a direct comparison between our experiment results and the results from the data.

We find that the experiment agrees qualitatively with observed user behavior on SE non-technical forums. For example, in Fig.~\ref{fig:PrChoose28AnswersVoteData}, the experiment's probability to choose an answer versus its position is almost exactly the same as the empirical data. This is surprising because answers are presumably ordered by their quality in the empirical data, while our experiment orders answers arbitrarily with artificial scores. Much of the probability to choose an answer in real data could be due to its order and not its quality. That said, when 8 answers are visible, there is a stronger preference to choose the top answer in the empirical data than the experiment, therefore our experiment's assumption of arbitrary answer order is not in as strong agreement with the data when there are enough answers visible. However, as noted previously, the vast majority of questions have few answers (where agreement is strongest), therefore, in most cases, answer position seems to be a bigger factor in answer popularity than answer quality.

\subsection{Discussion: A Good Answer is Hard To Find}

Our experimental model of Q\&A helps elucidate factors affecting user choices of the best answers to questions. We find that an answer's position plays an important role in this decision and is strongly enhanced by perceived popularity, information load, and the attention given to top answers. We see strong agreement between our experimental model and empirical data, which demonstrates the experiment's success: it captures many aspects of real Q\&A systems, despite differences in the populations, and the different motivations, of MT workers and SE users.

The empirical data can also distinguish position bias and other factors that contributed to answer popularity. We first discover from the natural experiment that moving an answer up by one position increases the probability that it is chosen by up to 30\% compared to when answers are ordered at random. The probability of moving an answer just one position higher compared to the {\it original position}, however, could be almost double the lower-position's probability depending on the number of answers visible. This is because the randomized position probability (a probability of around 0.35 when 15 answers are visible) is a mixture of the probability to choose an answer when it is listed first (a probability of around 0.45) and the probability fo choose an answer when it is listed last (which would presumably have to be around 0.25 if 0.35 is the mean value). Furthermore, and more troubling, the probability to choose an answer versus its position is very similar in the experiment, where answers were ordered arbitrarily, and the empirical data, where answers were presumably ordered based on their quality. This suggests that position bias, rather than quality, is a major factor contributing to answer popularity. 

{What cognitive mechanisms may affect position bias?} When controlling for score and attention, one hypothesis is that users have a trust bias, i.e., they believe answers are ordered by their quality and therefore trust top answers \cite{AttentionClick}. Alternately, they may pick answers that grab their attention~\cite{DDM3Objects}. Finally, the primacy effect, in which individuals prefer the first object they see, may play a role. The primacy effect applies to the temporal order of items, but we find that almost all workers scroll from the top to the bottom of a page, mousing over answers in turn, therefore top answers are typically the first ones seen.

In the random condition, the primacy effect probably plays a role in creating the position bias because it adequately explains why users prefer the answers they see first when 8 answers are visible, even though it cannot explain why users pick the last answer when 2 answers are visible. Attention plays a lesser role because attention seems to be flat regardless of answer position. We cannot completely rule out the trust bias but workers are not told how answers are ordered, and presumably will not assume a relationship between answer order and quality. Furthermore, it would seem strange that users would use this heuristic to prefer the last answer when 2 answers are visible and the first few answers when 8 are visible. 

In the social influence condition, trust bias could instead play a larger role. After controlling for position, the effect of scores is negligible. It therefore seems plausible that the relative position of scores contributes to position bias, meaning users prefer top answers because they trust they are of higher quality than lower ones. This is the trust bias where the assessment of quality is due to the heuristic of answer scores (popularity). It is also interesting that scores draw people's attention towards the top answers more than the bottom answers, therefore scores increase attention given to top answers, which is known to bias choice preferences \cite{DDM3Objects}. This is not the sole reason for position bias because the experiment often has a stronger position bias than the attention-based model would predict. Other attention-based null models, including picking the most-moused-over answer and choosing random answers that were moused over, produce similar results.

{How do we improve the wisdom of crowd in Q\&A systems?} If we first determine the quality of answers after controlling for position bias, we can list answers from highest-quality to lowest-quality. Clearly, the highest-quality answer will then be more visible because they are at the top of the webpage. In this way, position bias can improve user interactions by making good answers much easier to find.

This requires finding high-quality answers. One way to do so is to initially randomize the position of answers for each user and not show the answer scores, but still record the position-aggregated popularity of each answer. This reduces the correlations between users by reducing perceived popularity and position bias. We find in our experiments that position aggregation consistently finds the same answers to be popular regardless of the number of answers visible, presumably due to answer quality. {For example, the popularity of the oldest 2 answers (averaged over answer position) are highly correlated between conditions where 2 or 8 answers are visible (Spearman rank correlations are 0.74 \& 0.58 for the social influence and random conditions, respectively, with p-values $\le 0.01$).}

\section{Conclusion \& Future Work}

This paper analyzes factors that can affect the performance of Q\&A systems. First, we create a controlled experiment on MT that replicates the main functionality of SE by asking workers to choose the best answers to questions taken from an actual Q\&A forum. Controlling how, and in what order, the answers are shown to workers enables us to disentangle the effects contributing to their choices of best answers. We find that an answer's position strongly affects the probability that it will be chosen, and that this effect increases with cognitive load, perceived popularity, and attention. Perceived popularity alone, however, is not a significant factor. Next, we use empirical data and a natural experiment to elucidate the cognitive factors affecting answer popularity. Overall, we find broad agreement with our experiment results, which gives us confidence about its ecological validity. Because our results apply to different types of Q\&A boards on SE, we believe that they are widely applicable, at least among Q\&A systems. Furthermore, our observations are in line with recent work showing that position bias, when coupled with popularity-based ranking, reduces collective ability to identify highest-quality items~\cite{Abeliuk2017www}.

Although we do not find a correlation between the probability to choose an answer and perceived popularity when we correct for answer position, our current experiment does not completely decouple answer score from position, because answers are ranked from highest to lowest score. This motivates experiments where answer score and position are uncorrelated. Initial results from these experiments, however, agree with our current results: the effect of perceived popularity alone, especially when many answers are visible, is minimal. Future work will further explore these results.

Furthermore, although we find strong qualitative agreement between our model and empirical data, a common critique in any experiment is ecological validity: how alike is our experimental condition to the real world? For example, MT and SE users represent different populations, e.g., there may be more fluent English speaking workers in our experiment than in the board, which caters to individuals who are still learning English. Furthermore, workers have extrinsic motivations (getting paid), while SE users may be more intrinsically motivated, although the ``reputation'' and badges they receive for good answers and questions could also be interpreted as a form of extrinsic motivation. Our experimental design and choice of questions, focusing on general interest questions about the English language, minimizes this risk, but cannot rule it out completely. In order to address this potential critique, experiments should be created on MT in which workers sequentially upvote answers to questions (in the same way scores are created on SE). One can compare the popularity of answers versus their position in SE and MT using this method to determine whether both groups of people have the same motivations to choose answers. Furthermore, this experiment can help determine whether answer popularity through sequential voting agrees well with position-averaged answer popularity (a proxy for quality). Strong disagreement would add further evidence that answer score and answer quality are decoupled.

\subsection*{Acknowledgments}
Our work is supported by the Army Research Office under contract W911NF-15-1-0142 and by the Defense Advanced Research Projects Agency under contract W911NF-18-C-0011.


\begin{thebibliography}{}

\bibitem[\protect\citeauthoryear{Abeliuk \bgroup et al\mbox.\egroup
  }{2017}]{Abeliuk2017www}
Abeliuk, A.; Berbeglia, G.; Hentenryck, P.~V.; Hogg, T.; and Lerman, K.
\newblock 2017.
\newblock Taming the unpredictability of cultural markets with social
  influence.
\newblock In {\em Proceedings of the 26th International World Wide Web
  Conference (WWW2017)}.
\newblock Republic and Canton of Geneva, Switzerland: International World Wide
  Web Conferences Steering Committee.

\bibitem[\protect\citeauthoryear{Adamic \bgroup et al\mbox.\egroup
  }{2008}]{BestAnswerYahoo}
Adamic, L.~A.; Zhang, J.; Bakshy, E.; and Ackerman, M.~S.
\newblock 2008.
\newblock Knowledge sharing and yahoo answers: Everyone knows something.
\newblock In {\em Proceedings of the 17th international conference on World
  Wide Web},  665--674.
\newblock New York, NY: ACM.

\bibitem[\protect\citeauthoryear{Baron}{1986}]{InfoOverloadPsychology}
Baron, R.~S.
\newblock 1986.
\newblock Distraction-conflict theory: Progress and problems.
\newblock {\em Advances in experimental social psychology} 19:1--39.

\bibitem[\protect\citeauthoryear{Burghardt \bgroup et al\mbox.\egroup
  }{2017}]{MyopiaCrowd}
Burghardt, K.; Alsina, E.~F.; Girvan, M.; Rand, W.; and Lerman, K.
\newblock 2017.
\newblock The myopia of crowds: A study of collective evaluation on stack
  exchange.
\newblock {\em PLOS ONE} 12(3):e0173610.

\bibitem[\protect\citeauthoryear{Celis, Krafft, and Kobe}{2016}]{Celis2016}
Celis, L.~E.; Krafft, P.~M.; and Kobe, N.
\newblock 2016.
\newblock Sequential voting promotes collective discovery in social
  recommendation systems.
\newblock In {\em Proceedings of the Tenth International AAAI Conference on Web
  and Social Media (ICWSM 2016)},  42--51.
\newblock AAAI Press.

\bibitem[\protect\citeauthoryear{Chu, Anderson, and Sohn}{2001}]{MouseEye}
Chu, M.; Anderson, J.~R.; and Sohn, M.~H.
\newblock 2001.
\newblock What can a mouse cursor tell us more?: correlation of eye/mouse
  movements on web browsing.
\newblock In {\em CHI EA '01 CHI '01 Extended Abstracts on Human Factors in
  Computing Systems},  281--282.

\bibitem[\protect\citeauthoryear{Craswell \bgroup et al\mbox.\egroup
  }{2008}]{AttentionClick2}
Craswell, N.; Zoeter, O.; Taylor, M.; and Ramsey, B.
\newblock 2008.
\newblock An experimental comparison of click position-bias models.
\newblock In {\em WSDM '08 Proceedings of the 2008 International Conference on
  Web Search and Data Mining},  87--94.

\bibitem[\protect\citeauthoryear{de Condorcet}{1976}]{JuryThm}
de~Condorcet, M.
\newblock 1976.
\newblock {\em ``Essay on the Application of Mathematics to the Theory of
  Decision-Making.'' Reprinted in Condorcet: Selected Writings}.
\newblock Indianapolis, Indiana: Bobbs-Merrill,.

\bibitem[\protect\citeauthoryear{Ferrara \bgroup et al\mbox.\egroup
  }{2017}]{Ferrara17}
Ferrara, E.; Alipoufard, N.; Burghardt, K.; Gopal, C.; and Lerman, K.
\newblock 2017.
\newblock Dynamics of content quality in collaborative knowledge production.
\newblock In {\em ICWSM '17 Proceedings of the 11th International AAAI
  Conference on Web and Social Media}.

\bibitem[\protect\citeauthoryear{Galton}{1908}]{Galton1908}
Galton, F.
\newblock 1908.
\newblock Vox populi.
\newblock {\em Nature} 75:450--451.

\bibitem[\protect\citeauthoryear{Glenski, Johnston, and
  Weninger}{2015}]{SocialInfluenceBias2}
Glenski, M.; Johnston, T.~J.; and Weninger, T.
\newblock 2015.
\newblock Random voting effects in social-digital spaces: A case study of
  reddit post submissions.
\newblock In {\em Proceedings of the 26th ACM Conference on Hypertext \& Social
  Media},  293--297.
\newblock New York, NY: ACM.

\bibitem[\protect\citeauthoryear{Herbst and Mas}{2015}]{ObservationalVsExp}
Herbst, D., and Mas, A.
\newblock 2015.
\newblock Peer effects on worker output in the laboratory generalize to the
  field.
\newblock {\em Science} 350(6260):545--549.

\bibitem[\protect\citeauthoryear{Ho and Imai}{2006}]{BallotOrderCA}
Ho, D.~E., and Imai, K.
\newblock 2006.
\newblock Randomization inference with natural experiments.
\newblock {\em J. Amer. Statist. Assoc.} 101(475):888--900.

\bibitem[\protect\citeauthoryear{Ho and Imai}{2008}]{BallotOrderCA2}
Ho, D.~E., and Imai, K.
\newblock 2008.
\newblock Estimating causal effects of ballot order from a randomized natural
  experiment.
\newblock {\em Public Opinion Quarterly} 72(2):216--240.

\bibitem[\protect\citeauthoryear{Joachims \bgroup et al\mbox.\egroup
  }{2005}]{AttentionClick}
Joachims, T.; Granka, L.; Pan, B.; Hembrooke, H.; and Gay, G.
\newblock 2005.
\newblock Accurately interpreting clickthrough data as implicit feedback.
\newblock In {\em SIGIR '05 Proceedings of the 28th annual international ACM
  SIGIR conference on Research and development in information retrieval},
  154--161.

\bibitem[\protect\citeauthoryear{Kaniovski and Zaigraev}{2011}]{Kaniovski}
Kaniovski, S., and Zaigraev, A.
\newblock 2011.
\newblock Optimal jury design for homogeneous juries with correlated votes.
\newblock {\em Theory Dec.} 71:439--459.

\bibitem[\protect\citeauthoryear{Kittur and Kraut}{2008}]{WikiWisdomOfCrowds}
Kittur, A., and Kraut, R.~E.
\newblock 2008.
\newblock Harnessing the widom of crowds in wikipedia: Quality through
  coordination.
\newblock In {\em CSCW '08 Proceedings of the 2008 ACM conference on Computer
  supported cooperative work},  37--46.

\bibitem[\protect\citeauthoryear{Krafft \bgroup et al\mbox.\egroup
  }{2016}]{Krafft2016}
Krafft, P.~M.; Zheng, J.; Pan, W.; Penna, N.~D.; Altshuler, Y.; Shmueli, E.;
  Tenenbaum, J.~B.; and Pentland, A.
\newblock 2016.
\newblock Human collective intelligence as distributed bayesian inference.
\newblock {\em arXiv preprint:1608.01987}.

\bibitem[\protect\citeauthoryear{Krajbich and Rangel}{2011}]{DDM3Objects}
Krajbich, I., and Rangel, A.
\newblock 2011.
\newblock Multialternative drift-diffusion model predicts the relationship
  between visual fixations and choice in value-based decisions.
\newblock {\em PNAS} 108(33):13852--13857.

\bibitem[\protect\citeauthoryear{Krajbich, Armel, and
  Rangel}{2010}]{DDM2Objects}
Krajbich, I.; Armel, C.; and Rangel, A.
\newblock 2010.
\newblock Visual fixations and the computation and comparison of value in
  simple choice.
\newblock {\em Nature Neuroscience} 13(10).

\bibitem[\protect\citeauthoryear{Krumme \bgroup et al\mbox.\egroup
  }{2012}]{MusicLabModel}
Krumme, C.; Cebrian, M.; Pickard, G.; and Pentland, S.
\newblock 2012.
\newblock Quantifying social influence in an online cultural market.
\newblock {\em PLoS ONE} 7(5):e33785.

\bibitem[\protect\citeauthoryear{Lerman and Hogg}{2014}]{lerman14as}
Lerman, K., and Hogg, T.
\newblock 2014.
\newblock Leveraging position bias to improve peer recommendation.
\newblock {\em PLOS ONE} 9(6):e98914.

\bibitem[\protect\citeauthoryear{Lim and Van
  Der~Heide}{2015}]{lim2015evaluating}
Lim, Y.-s., and Van Der~Heide, B.
\newblock 2015.
\newblock Evaluating the wisdom of strangers: The perceived credibility of
  online consumer reviews on yelp.
\newblock {\em Journal of Computer-Mediated Communication} 20(1):67--82.

\bibitem[\protect\citeauthoryear{Lorenz \bgroup et al\mbox.\egroup
  }{2011}]{Lorenz2011}
Lorenz, J.; Rauhut, H.; Schweitzer, F.; and Helbing, D.
\newblock 2011.
\newblock {How social influence can undermine the wisdom of crowd effect.}
\newblock {\em Proceedings of the National Academy of Sciences}
  108(22):9020--9025.

\bibitem[\protect\citeauthoryear{Mantonakis \bgroup et al\mbox.\egroup
  }{2009}]{Primacy1}
Mantonakis, A.; Rodero, P.; Lesschaeve, I.; and Hastie, R.
\newblock 2009.
\newblock Order in choice: Effects of serial position on preferences.
\newblock {\em Psychol. Sci.} 20(11):1309--1312.

\bibitem[\protect\citeauthoryear{Muchnik, Aral, and
  Taylor}{2013}]{SocialInfluenceBias}
Muchnik, L.; Aral, S.; and Taylor, S.~J.
\newblock 2013.
\newblock Social influence bias: A randomized experiment.
\newblock {\em Science} 341:647--651.

\bibitem[\protect\citeauthoryear{Nematzadeh \bgroup et al\mbox.\egroup
  }{2016}]{TwitchInfoOverload}
Nematzadeh, A.; Ciampaglia, G.~L.; Ahn, Y.-Y.; and Flammini, A.
\newblock 2016.
\newblock Information overload in group communication: From conversation to
  cacophony in the twitch chat.
\newblock {\em arXiv:1610.06497}.

\bibitem[\protect\citeauthoryear{Oktay, Taylor, and Jensen}{2010}]{Oktay10}
Oktay, H.; Taylor, B.~J.; and Jensen, D.~D.
\newblock 2010.
\newblock Causal discovery in social media using quasi-experimental designs.
\newblock In {\em SOMA '10 Proceedings of the First Workshop on Social Media
  Analytics},  1--9.
\newblock ACM.

\bibitem[\protect\citeauthoryear{Ray}{2006}]{PredictMarket}
Ray, R.
\newblock 2006.
\newblock Prediction markets and the financial ``wisdom of crowds''.
\newblock {\em Journal of Behavioral Finance} 7(1):2--4.

\bibitem[\protect\citeauthoryear{Salganik, Dodds, and Watts}{2006}]{MusicLab}
Salganik, M.; Dodds, P.; and Watts, D.
\newblock 2006.
\newblock Experimental study of inequality and unpredictability in an
  artificial cultural market.
\newblock {\em Science} 311:854--856.

\bibitem[\protect\citeauthoryear{Simpson}{1951}]{SimpsonEffect}
Simpson, E.
\newblock 1951.
\newblock The interpretation of interaction in contingency tables.
\newblock {\em J. R. Stat. Soc.} 13:238--241.

\bibitem[\protect\citeauthoryear{Stoddard}{2015}]{PopularDynam}
Stoddard, G.
\newblock 2015.
\newblock Popularity dynamics and intrinsic quality in reddit and hacker news.
\newblock In {\em Proceedings of the Ninth International AAAI Conference on Web
  and Social Media},  416--425.

\bibitem[\protect\citeauthoryear{Surowiecki}{2005}]{surowiecki2005wisdom}
Surowiecki, J.
\newblock 2005.
\newblock {\em The wisdom of crowds}.
\newblock New York: Anchor.

\bibitem[\protect\citeauthoryear{Yao \bgroup et al\mbox.\egroup
  }{2015}]{HighQualityQA}
Yao, Y.; Tong, H.; Xie, T.; Akoglu, L.; Xu, F.; and Lu, J.
\newblock 2015.
\newblock Detecting high-quality posts in community question answering sites.
\newblock {\em Information Sciences} 302:70--82.

\bibitem[\protect\citeauthoryear{Yucesoy and
  Barab{\'a}si}{2016}]{PerformanceSuccess}
Yucesoy, B., and Barab{\'a}si, A.-L.
\newblock 2016.
\newblock Untangling performance from success.
\newblock {\em EPJ Data Science} 5(1):1.

\end{thebibliography}

\end{document}